\documentclass[journal, twocolumn, 10pt]{IEEEtran}
\usepackage{graphicx}
\usepackage{algorithm}
\usepackage{algorithmicx}
\usepackage{algpseudocode}
\usepackage{cite}
\usepackage{color}
\usepackage{amsmath,amssymb,amsfonts,amsthm,dsfont}
\usepackage{bm}
\usepackage{setspace}
\usepackage{multirow}
\usepackage{mathtools}
\usepackage{epstopdf}
\usepackage{url}

\begin{document}

\title{Traffic-aware Patching for Cyber Security \\ in Mobile IoT}


\author{
    \IEEEauthorblockN{Shin-Ming Cheng\IEEEauthorrefmark{1}, Pin-Yu Chen\IEEEauthorrefmark{2}, Ching-Chao Lin\IEEEauthorrefmark{1}, and Hsu-Chun Hsiao\IEEEauthorrefmark{3} 
 } \\
\IEEEauthorblockA{\IEEEauthorrefmark{1}Department of Computer Science and Information Engineering, National Taiwan University of Science and Technology, Taipei, Taiwan
\\ \{smcheng, m10415008\}@mail.ntust.edu.tw} \\
\IEEEauthorblockA{\IEEEauthorrefmark{2}IBM Thomas J. Watson Research Center, Yorktown Heights, New York, USA
\\ pin-yu.chen@ibm.com} \\
\IEEEauthorblockA{\IEEEauthorrefmark{3}Department of Computer Science and Information Engineering, National Taiwan University, Taipei, Taiwan
\\ hchsiao@csie.ntu.edu.tw} \\
}

\maketitle

\begin{abstract}
The various types of communication technologies and mobility features in Internet of Things (IoT) on the one hand enable fruitful and attractive applications, but on the other hand facilitates malware propagation, thereby raising new challenges on handling IoT-empowered malware for cyber security. Comparing with the malware propagation control scheme in traditional wireless networks where nodes can be directly repaired and secured, in IoT, compromised end devices are difficult to be patched. Alternatively, blocking malware via patching intermediate nodes turns out to be a more feasible and practical solution. Specifically, patching intermediate nodes can effectively prevent the proliferation of malware propagation by securing infrastructure links and limiting malware propagation to local device-to-device dissemination. This article proposes a novel traffic-aware patching scheme to select important intermediate nodes to patch, which applies to the IoT system with limited patching resources and response time constraint. Experiments on real-world trace datasets in IoT networks are conducted to demonstrate the advantage of the proposed traffic-aware patching scheme in alleviating malware propagation.
\end{abstract}

\IEEEpeerreviewmaketitle
\begin{IEEEkeywords}
heterogeneous links, IoT malware, patching
\end{IEEEkeywords}

\section{Introduction}
\label{sec_intro}

By integrating the ability of sensing physical world and the privilege in availing communication capabilities, Internet of Things (IoT) enables close interactions between humans and machines. IoT generally consists of numerous end IoT devices for sensing and action, intermediate nodes with wired connectivity for data relaying, and application servers in the cloud for data controlling and analysis. Typically, IoT devices can communicate with each other with minimal human intervention and build an autonomous and complex network. As the boundary between machines and humans gets blurry, adversaries in the cyberspace can threaten human users’ safety and privacy in the physical world. Obviously, the growing popularity of devices with rich wireless communication capabilities has made IoT attractive to digital viruses and malicious contents. Consequently, in recent years the security issues in IoT has been an ever-increasing concern~\cite{Granjal15,Minn15,Cheng16}.

From an adversary's perspective, the unique features of IoT facilitate the exploits of devices as well as the propagation of IoT malware. These features include constrained resources, heterogeneous links, and vulnerable usability, which are discussed as follows.

\begin{figure}[t]
    \centering
    \includegraphics[width=3.5in]{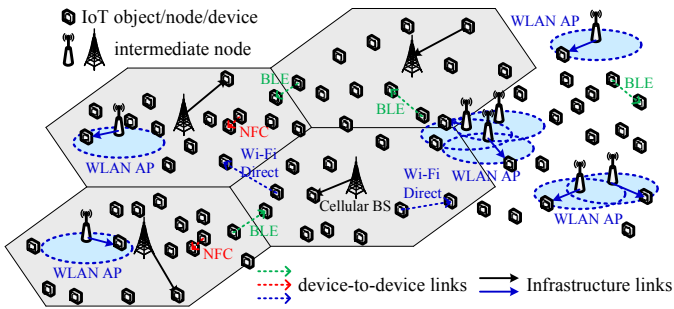}
    \caption{IoT platform with infrastructure and device-to-device links}
    \label{fig_IoTarch}
\end{figure}

\begin{LaTeXdescription}
\item [Resource-constrained IoT devices.] Comparing with the intermediate nodes located at the end side of the infrastructure with wired connectivity, IoT devices designed to perform simple sensing and actuation operations have limited computation and communication capabilities. In this case, the algorithm and mechanism applied on IoT devices are relatively simple. As a result, the attacker can spend much less resource to break in IoT devices, rendering them the targets of malicious users. For example, due to the overhead of certificate management and public-key cryptography, many existing IoT devices fail to support state-of-the-art secure communication protocols (e.g., SSL/TLS). Therefore, the adversary can eavesdrop on sensitive sensor data and even manipulate data without being detected. Another example is that IoT devices often have limited entropy sources, which results in weak cryptographic keys that can be predicted by the attacker. Moreover, since most IoT devices run on embedded Linux OS, the attacker can easily create IoT malware by recompiling existing Linux malware for other instruction set architectures.

\item [Heterogeneity.] In order to support different kinds of IoT applications, IoT devices are often equipped with heterogeneous communication and computation capabilities for the purpose of seamless operations. However, the heterogeneity and potentially vast amount of IoT devices facilitate the fabrication of identity and hiding of malware. Moreover, as shown in Fig.~\ref{fig_IoTarch}, compromised IoT devices might disseminate malware via heterogeneous communication links as described below.
\begin{itemize}
\item \textbf{Infrastructure links.} IoT malware can propagate using infrastructure-based communication technologies, such as GSM/GPRS/UMTS/LTE and WLAN, via intermediate nodes, such as access point (AP), base station (BS), or gateway. In particular, IoT malware inherits the threats caused by computer malware. Similar to computer malware, most IoT malware families today scan the IP address space for vulnerable victims and spread via the Internet. Due to the widespread use of weak login credentials and the fact that many IoT devices are Internet-accessible, some botnets have allegedly harvested more than one million of infected IoT devices.\footnote{\url{http://thehackernews.com/2016/10/iot-dyn-ddos-attack.html}}

\item \textbf{Device-to-device links.} IoT malware could exploit the proximity-based wireless media such as BLE, Wi-Fi Direct, and NFC to infect the devices in the vicinity~\cite{Zyba09}. In this case, IoT malware is stored and forwarded by taking advantages of mobility and ubiquity. For example, Colin O’Flynn in Black Hat USA 2016 as well as Ronen and Shamir~\cite{Ronen16} discussed the possibility of light bulb worm, which allows a reprogrammed bulb to re-flash nearby bulbs. 
\end{itemize}

\item [Usability.] Security is only as strong as its weakest link, and the weakest link, in many cases, is the humans who implement, operate, and use the system. For example, a proven secure cryptographic primitive, if implemented or used incorrectly, can still be circumvented. Moreover, users may choose to ignore or even bypass a security mechanism if it prevents (e.g., due to slow performance, badly designed user interface, and unclear instructions) the users from doing what they meant to do. Since IoT devices often lack convenient input and output interfaces, the original security features might be bypassed by the non-professional IT users, thereby increasing the possibility and risk of human errors and facilitating the spreading of malware~\cite{Minn15}.
\end{LaTeXdescription}

Obviously, the software updates and patching are necessary to prevent the IoT devices from being compromised. A single software flaw will make a tremendous range of IoT devices vulnerable to attacks since software components are reused in different devices.\footnote{http://blog.senr.io/blog/400000-publicly-available-iot-devices-vulnerable-to-single-flaw} However, without a friendly interface to get alerted about security updates, most users forget to update software installed in IoT devices and leave them out-of-date. In addition, without basic programming knowledge and security awareness, users might be unwilling to perform manual-download-and-install approach for the software update. As a result, it is critical to design a reasonable solution to prevent the occurrence of the large-scale malware propagation among trillion of unpatched, insecure, and even compromised IoT devices.

Instead of patching resource-constrained and UI-unfriendly compromised IoT devices directly, this article introduces a more feasible solution, where operators could only patch or recover IoT devices via infrastructure, i.e., securing the intermediate nodes. In this case, the patched AP, BS or gateway could stop the malware propagation by patching via infrastructure links. The concept of leveraging intermediate nodes to improve IoT security has appeared in the recent commercial product F-Secure SENSE.\footnote{https://community.f-secure.com/t5/F-Secure-SENSE/What-are-the-current-protection/ta-p/82972} However, its main purpose is to block malicious websites and IoT botnet masters instead of considering securing important infrastructure links between IoT devices and intermediate nodes. On the other hand, the idea behind IoT Sentinel~\cite{Miettinen16} is similar to our solution, where the type of IoT devices are identified by intermediate nodes, and the communications of vulnerable IoT devices are constrained by enabling enforcement of rules. Different from our solution, software-defined network (SDN) is exploited in IoT Sentinel for network flow isolation and for prevention of malware propagation. 


With limited efforts and resources, the operator might not be able to patch all intermediate nodes but only a portion of them. One naive method is to simply patch those intermediate nodes in a random order. However, a smarter approach is to protect the most important node first, as suggested by the framework of network robustness analysis~\cite{CPY15_PRE,CPY14_ComMag}. This article proposes a traffic-aware patching scheme, where the operator patches the intermediate nodes sequentially in a descending importance order. In particular, an intermediate node who could contact with a large number of IoT devices will be protected first. Moreover, such volume-based patching approach is effective to the current infamous DDoS attacks launched by IoT bots.

By leveraging a real-world trace datasets containing communication history over device-to-device and infrastructure links, we conduct an extensive experiment to demonstrate the effect on constraining malware propagation via infrastructure links. To the best of authors' knowledge, this article is the first work discussing the control of malware propagation from the perspective of infection paths, which could avail the damage estimation caused by the malware and improve the development of attack detection methods for IoT networks.


\section{How to compromise IoT devices?}
IoT devices are an attractive attack target for cybercriminals: IoT devices often employ weak security measures, and their compromise can lead to privacy breaches and safety threats in the real world. The insecurity of existing IoT devices has been highlighted repeatedly by security researchers and practitioners. Recently, several malware families were found to target vulnerable IoT devices (e.g., routers, IP cameras, and CCTVs) and form botnets for DDoS. It is estimated that some IoT botnets comprise more than one million of infected devices, and thus can generate high-volume DDoS traffic even without amplification. For example, in September 2016, an IoT botnet called Mirai crippled a website with 620 Gbps of attack traffic, which is almost twice as much as the biggest DDoS attack witnessed in 2015. Later in October 2016, the same botnet attacked the Dyn DNS service provider, taking down a large portion of websites in the North America, including GitHub, Twitter, Netflix, etc.\footnote{\url{https://www.us-cert.gov/ncas/alerts/TA16-288A}}  At DEF CON 2016, security researchers showed a proof-of-concept IoT ransomware that demands ransom for a hacked smart thermostat, which will be set to a high temperature without a timely payment.\footnote{https://www.pentestpartners.com/blog/thermostat-ransomware-a-lesson-in-iot-security/} As attackers are finding creative ways to monetize infected IoT devices, it is inevitable to see an increase of new IoT malware families that are more destructive and contagious than ever.

IoT malware can propagate via infrastructure links and/or device-to-device links. We discuss both cases in this section.

\subsection{Compromising IoT devices via infrastructure links}
Many of the IoT malware families today propagate via infrastructure links, particularly the Internet. Moreover, they share a common infection and spreading pattern: The attacker harvests new vulnerable IoT devices through address space scanning. This scanning can be performed by external servers, such as the C\&C servers, or by the compromised devices. The attacker targets Telnet- or SSH-accessible devices that use default or weak login credentials and thus can easily obtain root access permission by brute-force password cracking. Once the attacker gets the shell of the hacked device, malware payload will be downloaded and installed. IoTPOT~\cite{Minn15}, an IoT honeypot project, observed at least four IoT malware families that can propagate via Telnet. In addition to cracking weak passwords, some malware also exploits software vulnerabilities. For example, CCTV-targeting RADIATION malware exploits ShellShock and some known CCTV vulnerabilities to spread from device to device.



\subsection{Compromising IoT devices via device-to-device links}

Malware can also propagate in proximity via device-to-device links in additional to infrastructure links. Cabir and Commwarror are examples of mobile worms that spread via Bluetooth and infect mobile phones running Symbian OS.

Although we have not witnessed device-to-device IoT malware in the wild, it is theoretically possible. For example, researchers pointed out the possibility of light bulb worms that spread to nearby bulbs via Zigbee~\cite{Ronen16} and worms that infect wearable trackers and then spread to others by Bluetooth.\footnote{\url{http://www.theregister.co.uk/2015/10/21/fitbit_hack/}} Moreover, since proximity-based wireless interfaces are often always-on and users have no control to disable them, it would be difficult to contain malware propagation given the large attack surface.

Regardless of how malware propagates, the risk of self-replicating IoT malware is amplified by unpatched IoTs. Patching vulnerable IoT devices nevertheless remains extremely expensive and far from successful in practice. In 2015, Charlie Miller and Chris Valasek demonstrated remote exploitation of a Jeep, which forced Chrysler to recall and patch 1.4 million vehicles.\footnote{\url{https://www.wired.com/2015/07/hackers-remotely-kill-jeep-highway/}} Cui and Stolfo~\cite{Cui10} discovered more than 540,000 publicly accessible devices using default root passwords---an old yet persisting vulnerability since the invention of password-based authentication. Worse yet, the problems encountered when patching computers and mobile phones (e.g., privacy, legacy devices, and lack of incentives) will linger and even exacerbate when attempting to patch IoT devices.

\begin{figure*}[t]
    \centering
    \includegraphics[width=6in]{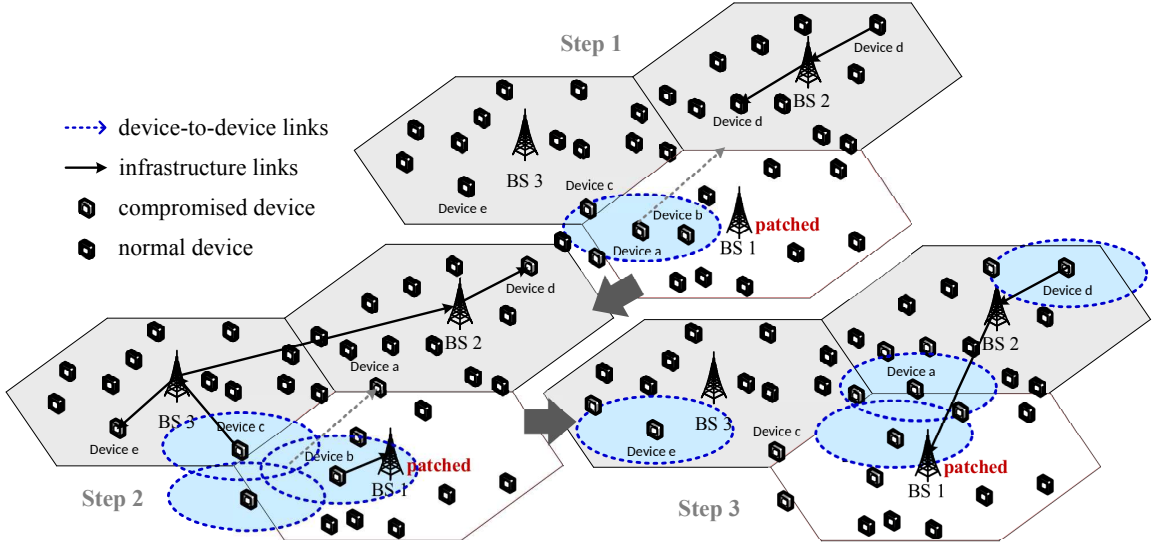}
    \caption{Illustration of malware propagation under the infrastructure patch scheme.}
    \label{fig_propagation}
\end{figure*}

\section{Modeling of IoT malware}

The topic of modeling malware/virus spreading has been investigated in a traditional scenario where computers or laptops are not connected to the Internet. Since the spread of epidemics among people is similar to the spread of malware over networks, the current literature adopts the idea from epidemiological models to build the models for malware on the assumption of homogeneous infection path~\cite{Peng14}. In the mobile environment, malware can propagate via intermittently connected networks by taking advantage of opportunistic encounters \cite{Tanachaiwiwat09}. Wang \textit{et al.}~\cite{Wang09} study spreading patterns of mobile phone viruses which may traverse through multimedia messaging services (MMS) or Bluetooth by simulations. Cheng and Chen~\cite{Cheng11} further models malware propagation in generalized social networks consisting of delocalized and localized links.  

From the discussion of the previous section, we understand that in practice patching the compromised IoT devices is difficult to be achieved. Consequently, the current formulation of malware propagation and the control model~\cite{CPY14_Cybernetics} can not be applied directly in the IoT field. Typically, in one of the most famous susceptible-infection-recover (SIR) model, the malware is assumed to be detected and repaired at each node, which reflects the transition from ``infected'' state to ``recovered'' state.  Regarding the IoT device who detects the malware, instead of directly patching it, it is more feasible to patch on the infrastructure side to prevent further spreading of malware. In this case, compromised IoT devices located in the coverage area of the patched intermediate nodes are controlled, that is, malware cannot be propagated via patched intermediate nodes. As a result, the infrastructure links can be regarded as ``recovered'' while the compromised IoT device remains ``infected'' using the terminology of SIR model. The observation that malware control in IoT environment can be cast as a ``link recovery'' problem instead of a ``node recovery'' problem motivates a different development of modeling and formulation.

\section{Feasible Patching Schemes in IoT Environment} 

This section proposes patching schemes for IoT environment, where we can only control infrastructure links but not the compromised nodes themselves. The patching scheme consists of several phases. In the \textit{detecting phase}, infrastructure leverages traditional IDS or firewall to identify the existence of malware or compromised node. Once a malicious code is found to be propagated from the compromised IoT devices, \textit{patching phase} starts to analyze the malware and patches the intermediate nodes according to patching sequence to prevent the large-scale propagation of malware. In practice, intermediate nodes are capable of performing resource intensive tasks and thus can support over the air (OTA) update mechanisms. In the \textit{patching phase}, such OTA mechanisms allow the administrator to remotely install required update on the intermediate nodes, thereby ensuring timely mitigation of compromised nodes. In addition, since intermediate nodes are significantly fewer than IoT devices, the administrator can also manually patch legacy intermediate nodes that do not support OTA update.

Fig.~\ref{fig_propagation} describes an example of how a compromised device propagates malware in IoT environment with patched and unpatched intermediate nodes. For the devices located in the coverage area of the patched intermediate nodes, two possible operations will be executed. 
\begin{itemize}
\item \textbf{Compromised devices} can distribute malware via device-to-device links but not infrastructure links. As shown in step 1 of Fig.~\ref{fig_propagation}, the compromised device propagates malware to devices \textit{b} and \textit{c} in the vicinity. However, in step 2 of Fig.~\ref{fig_propagation}, device \textit{b} cannot propagate malware via infrastructure link since the malware is blocked at the patched BS. 
\item \textbf{Normal devices} can only be compromised via device-to-device links since the malware propagated from infrastructure will be identified and blocked by the patched intermediate nodes. For example, in step 3 of Fig.~\ref{fig_propagation}, device \textit{d} propagates malware from BS 2 to BS 1, however, the patched BS 1 will not relay the malware to any device in its coverage area.  
\end{itemize}
For the devices located in the coverage area of the unpatched intermediate nodes, there are no means to prevent malware propagation. For example, in step 2 of Fig.~\ref{fig_propagation}, device \textit{c} under unpatched BS 3 could infect device \textit{d} controlled by unpatched BS 2 via infrastructure links. Moreover, device \textit{a} moving from patched BS 1 to unpatched BS 2 could propagate malware via device-to-device links freely.

\begin{algorithm}[h]
    \caption{Traffic-aware Patching} \label{alg_traffic}
        \begin{algorithmic}[1]
        \Require{The set of intermediate nodes, $S_{AP}$; The time to start patching, $t_{p}$; The percentage of patched intermediate nodes, $p$} 	
        \If {$currentTime < t_{p}$}
        	\State Collect traffic information for each intermediate node
        \Else
        	\If {$currentTime >= t_{p}$}
        	    \State Sort intermediate nodes according to the 
        	    \State ~~~importance metric in descending traffic order
        		\State Patch top $p\%$ $S_{AP}$
        	\EndIf
        \EndIf
        \end{algorithmic}
\end{algorithm}

Algorithm~\ref{alg_traffic} describes the detailed steps in \textit{patching phase}. With limited resources and efforts, the operator could provide a fixed amount of patches on the intermediate nodes (e.g., $p$ percentage). To alleviate the propagation from the infrastructure links, the $p \%$ most important intermediate nodes will be chosen for patching. It is similar to the idea of protecting the most important node to maintain network robustness~\cite{CPY15_PRE}. As a result, we introduce the traffic monitoring duration (see lines 1-2 in Algorithm~\ref{alg_traffic}) for evaluating the importance of intermediate nodes. From the monitored results, the proposed traffic-aware patching scheme sorts the intermediate nodes in descending order according to the traffic volumes (see lines 5-6 in Algorithm~\ref{alg_traffic}), and the top $p\%$ intermediate nodes are patched (see line 7 in Algorithm~\ref{alg_traffic}).


Obvious, the proposed volume-based patching is effective to the attack which generates a large number of traffic volume, e.g., DDoS attacks. The patched intermediate nodes could prevent the redirection of malicious traffic introduced by the DDoS attack launched by the IoT botnets.

\section{Performance Evaluation}
\label{sec_sim}

In this section, we implement the proposed traffic-aware patching scheme and compare its performance with a randomized patching scheme on real-life traffic traces collected from a mobile social network consisting of 59 users (devices) and 1751 APs~\cite{Dong11}. In this network, each user can communicate with other users through two types of links: (1) an infrastructure link via (possibly multiple) APs, and (2) a direct device-to-device link to users within transmission range. These two types of links among users are similar to the illustration of mobile IoT in Fig. \ref{fig_IoTarch}. As we mentioned in previous sections, in this experiment infrastructure links can be made secure via patching, whereas direct device-to-device links are vulnerable to potential security threats.

Following the vulnerability analysis of transmissive attacks in~\cite{Cheng16}, we simulate the propagation dynamics of self-replication malicious codes by first randomly selecting a user in the network as the initially compromised device. Then, using the actual traces of communication patterns provided by the dataset~\cite{Dong11}, each infected device can compromise its contact through an infrastructure link with probability $\lambda_{\textnormal{inf}}$, and can compromise its contact through a direct device-to-device link  with probability $\lambda_{\textnormal{dir}}$. Specifically, if one of the APs in the communication path between one infected device to its contact has been successfully patched, then the malware propagation is in vain due to enhanced security.

For traffic-aware patching,  we are interested in investigating the trade-offs between the time spent in analyzing traffic volumes (i.e., the traffic monitoring duration) and the time instance to patch APs (i.e., the patch time). As described in the previous section, given a fixed amount of patches, the proposed traffic-aware patching scheme sorts the APs in descending order according to the traffic volumes in the traffic monitoring duration, and provides patches to the top APs. Intuitively, longer traffic monitoring duration better specifies the important APs in communicating devices. However, longer traffic monitoring duration also leads to more exploits in security vulnerabilities due to later patch time. As a result, given a fixed amount of patches, we aim to study the non-trivial optimal patch time that collects sufficient traffic information for patching while minimizing the security risks. 

\begin{figure*}[!t]
    \centering
    \includegraphics[width=6in]{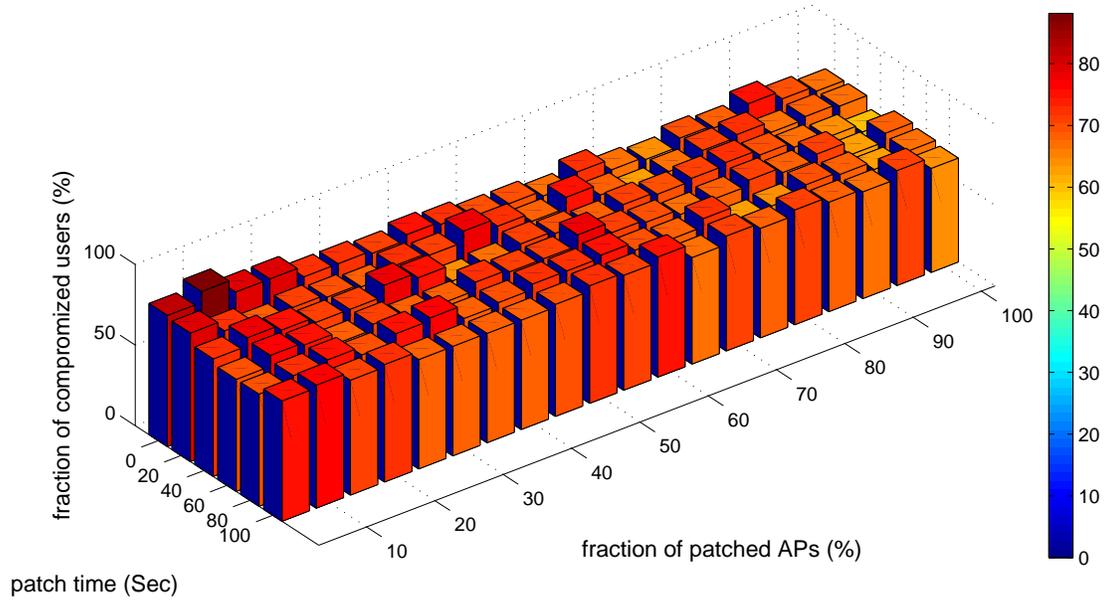}
    \caption{Fraction of compromised users with respect to different patch time and patched APs under the traffic-aware patching scheme. $\lambda_{\textnormal{inf}}=0.00004$ and $\lambda_{\textnormal{dir}}=0.00001$. The results are averaged over 500 trials.} \label{fig_degree}
\end{figure*}

\begin{figure*}[!t]
	\centering
	\includegraphics[width=6in]{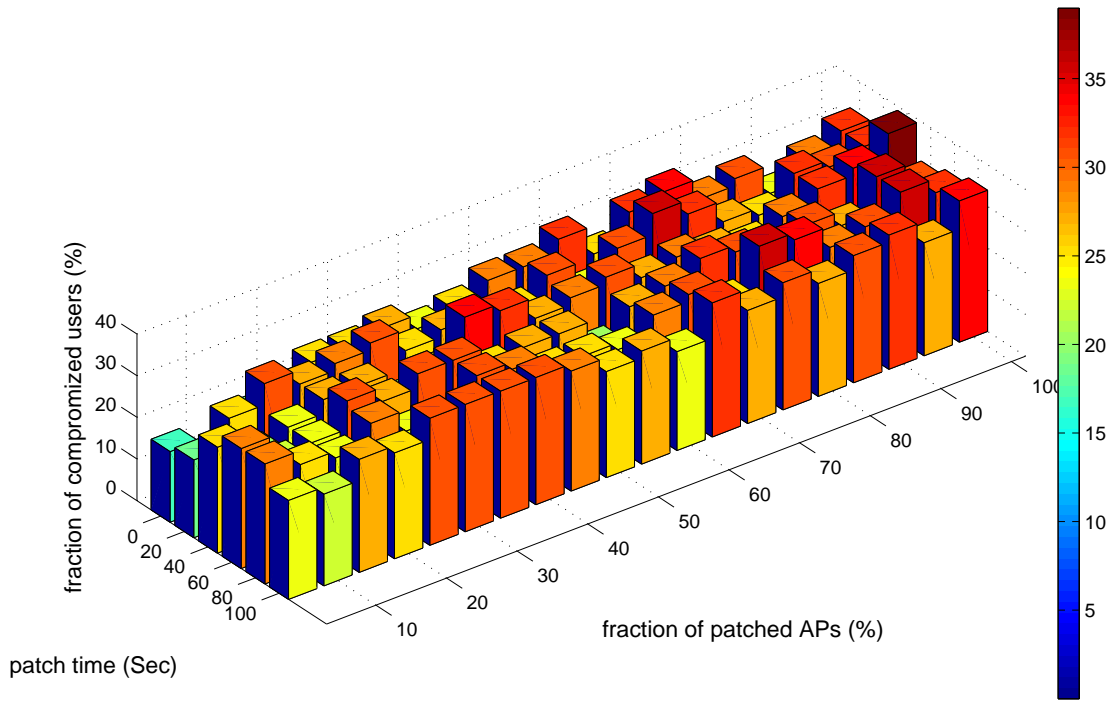}
	\caption{Performance comparison of traffic-aware patching scheme versus the no-patching scheme. This figures shows the difference of compromised users between the no-patching scheme and the traffic-aware scheme.  $\lambda_{\textnormal{inf}}=0.00004$ and $\lambda_{\textnormal{dir}}=0.00001$. The results are averaged over 500 trials.} \label{fig_degree_2}
\end{figure*}

\begin{figure}[!t]
    \centering
    \includegraphics[width=3.2in]{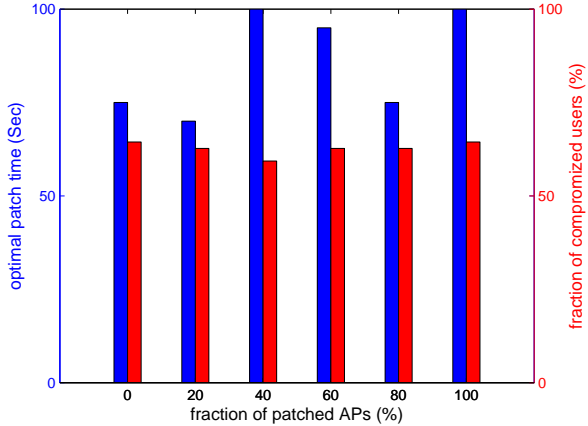}
    \caption{Optimal patch time and the corresponding number of compromised users given patched APs. $\lambda_{\textnormal{inf}}=0.00004$ and $\lambda_{\textnormal{dir}}=0.00001$. The results are averaged over 500 trials.} \label{fig_optimal}
\end{figure}

\begin{figure*}[!t]
	\centering
	\includegraphics[width=6in]{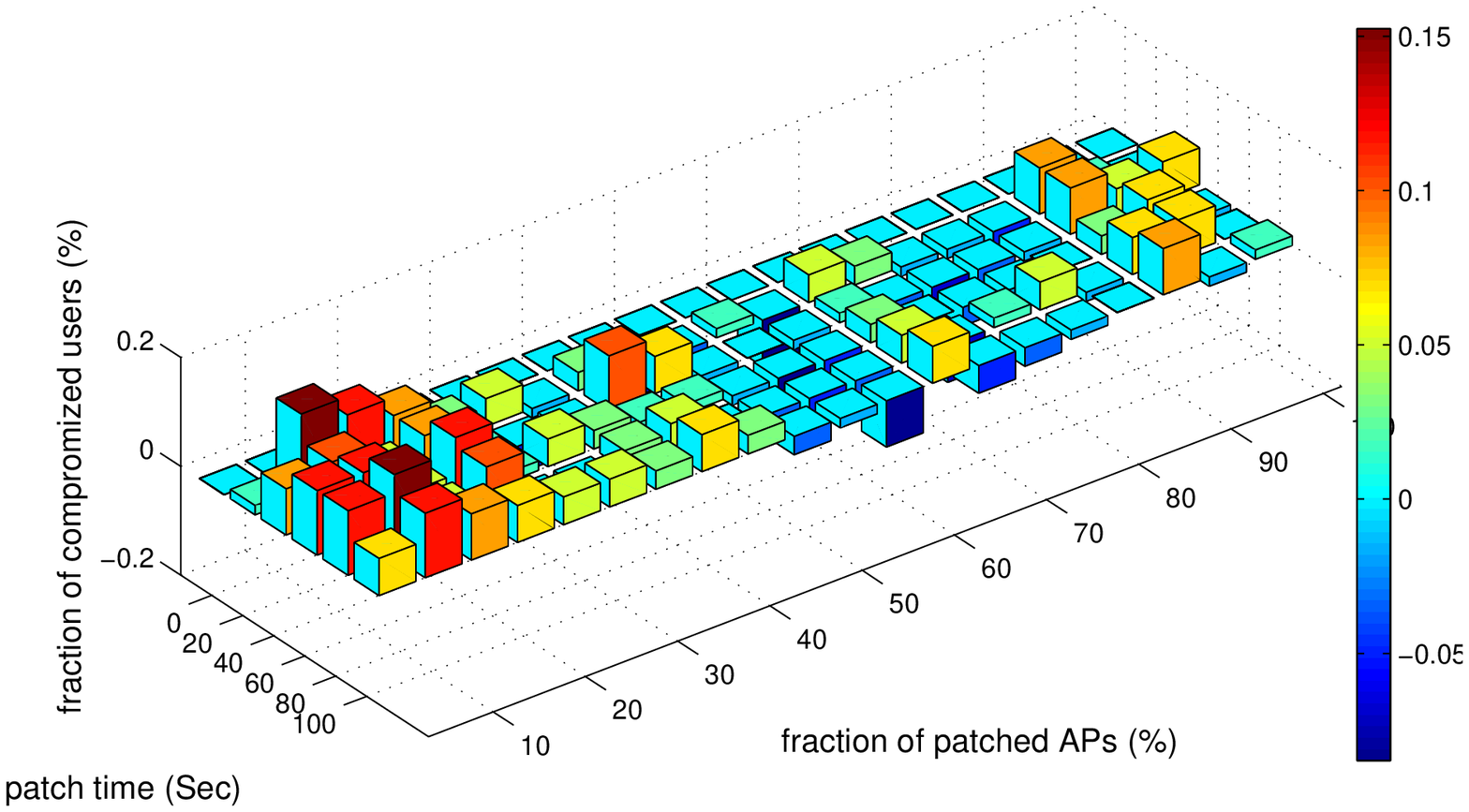}
	\caption{Performance comparison between random patching and traffic-aware patching in terms of the difference between the fraction of compromised users under random patching to that of traffic-aware patching. $\lambda_{\textnormal{inf}}=0.00004$ and $\lambda_{\textnormal{dir}}=0.00001$. The results are averaged over 500 trials.} 
	\label{fig_difference}
\end{figure*}

Fig.~\ref{fig_degree} shows the fraction of compromised users with respect to different patch time and patched APs under the traffic-aware patching scheme. To demonstrate the effectiveness of the proposed traffic-aware patching scheme, Fig. \ref{fig_degree_2} further compares the difference of compromised users between the no-patching scheme and the traffic-aware scheme. It can be observed that the best patching strategy that leads to a maximal decrease in the number of compromised users compared with the no-patching scheme is to monitor the traffics for 40 seconds and then provide patches to all APs. Note that 100\% patched APs (i.e., securing all infrastructure links) with patch time 0 may not be the optimal patch strategy since the malicious codes are still able to propagate through direct device-to-decide links. To further understand the effect of traffic-aware patching, for a given fraction of patched APs, Fig.~\ref{fig_optimal} shows the optimal patch time that leads to the lowest total number of compromised users. We observe that if one is able to patch more APs, then late patch time can have better performance, which suggests that traffic volumes are indeed important information for patching.

For fair comparison, we also compare the performance of traffic-aware patching with random patching. Random patching provides immediate patches (i.e., has patch time $0$ ) and randomly selects a fraction of APs to patch. Fig. \ref{fig_difference} shows the difference between the fraction of compromised users under random patching to that of traffic-aware patching, where larger positive values imply traffic-aware patching is more effective in securing the network, and vice versa. We observe that traffic-aware patching is significantly better than random patching in the regime of few patched APs (e.g., below 30\%). Moreover, given a fixed fraction of patched APs, for traffic-aware patching, there is at least one patch time that leads to either better or identical performance compared with random patching, which suggests the robustness and reliability of the proposed patching scheme. Even in the regime of many patched APs (e.g., above 90\%), the performance of traffic-aware patching is still superior to random patching, which suggests the importance of patching APs of high traffic volumes for enhanced security.



\section{Some Ongoing Challenges and Open Research Questions}
Here we discuss several ongoing challenges and open research questions related to IoT malware propagation and patching.

\begin{itemize}

\item \textbf{Transfer learning for optimal patch time.}~\\
In the experiments, we find that the patch time is crucial to preventing malware propagation. How to design and simulate realistic testbeds to assist determining the optimal patch time and to enable transfer learning for defending real-life unknown security threats are ongoing challenges.

\item \textbf{Predictive malware propagation models for mobile IoT.} ~\\
In this article, we have addressed patching issues in mobile IoT as link recovery instead of node recovery, where the latter has been extensively studied in traditional wireless networking scenarios. How to establish effective mathematical models for predicting malware propagation dynamics in mobile IoT that take into account the traffic-aware and random patching schemes are new research challenges.

\item \textbf{Various importance metrics for intermediate nodes.}
The proposed scheme simply applies traffic volume as the metric to determine the importance of intermediate nodes and the patching sequence. It can be regarded as protecting the entire network by patching a relatively small fraction of intermediate nodes with the highest ``degree'' metric. The operator could consider more information about intermediate nodes, such as the topology of intermediate nodes, in order to design a more effective importance metric for determining the patching sequence. For example, the ``betweenness'' metric could be leveraged, which is defined as the fraction of all shortest paths passing through the node among all shortest paths between each node pair in the network. 

\item \textbf{Patching via path-based traffic patterns.}
The proposed traffic-aware patching scheme only considers the one-hop traffic information in terms of the traffic volume from IoT devices to intermediate nodes. The patching scheme could benefit from the knowledge beyond one-hop information, such as the path-based end-to-end traffic patterns. However, path-based traffic patterns are relatively difficult to be collected or acquired compared to the one-hop traffic information.

\item \textbf{How to achieve (virtually) patch?}
IoT devices often lack friendly user interfaces and are left unattended after installation. As a consequence, users have trouble knowing whether a device is hacked, and even they do, they may find it challenging to \textit{manually patch} the device: they need to retrieve updated firmware online, access the hacked device, install the firmware, etc. Thus, \textit{automatic patching} is needed to secure IoT at scale.

One promising direction is for IoT devices to support Firmware Over The Air (FOTA), as most PCs and mobile phones do nowadays. However, an efficient and secure FOTA for IoT remains an open challenge due to the heterogeneity of IoT networks. For example, transport security and code signing are required to ensure the authenticity of the updated firmware. The IoT gateway might help reduce the overhead by caching and offloading the security check. Moreover, the human factors need to be taken into consideration as well. As in the PC and mobile phone worlds, forcing software update without explicit user consent can be disastrous. It can even be life-threatening if the update happens at a wrong time (e.g., updating the vehicle while driving).
\end{itemize}

\section{Concluding Remarks}
This article considers the security threats incurred by the heterogeneous links of IoT and designs a novel patching scheme to alleviate malware propagation. Instead of the impractical solution of directly patching compromised IoT devices, we propose to patch important intermediate nodes  based on the traffic volumes to prevent major security exploits and to avoid catastrophic malware propagation. With the proposed traffic-aware patching scheme, malware propagation is restricted to direct device-to-device connection, and therefore the damage of malware propagation can be significantly reduced. We conduct experiments in IoT environment to demonstrate the effectiveness of the proposed traffic-aware patching scheme, and we also discuss some ongoing research challenges and open research questions related to IoT patching.

The proposed traffic-aware patching scheme and the experimental results bring new insights to IoT security. For instance, the infeasibility of direct patching on IoT devices calls for new IoT malware models and security assessment approaches. The experimental results can assist in developing new attack detection techniques and patching strategies for preventing malware propagation.
Obviously, the resource-constrained, user-unfriendly, and heterogeneous features of IoT devices hinder the security design and development for IoT. However, the experimental results indicate a promising method to secure the entire IoT system by patching intermediate nodes. In summary, we provide two guidelines for how to consider cyber security when designing IoT systems accordingly:
\begin{itemize}
\item The consideration of intermediate nodes that bridge the gap between resource-constrained IoT devices and powerful IoT application servers is necessary when designing cyber security for IoT. By shifting computation-consuming, security related functionalities (e.g., flow identification, filtering, and isolation) to intermediate nodes, they can play the roles of the onsite guards. In particular, the flexibility and reconfigurability of intermediate nodes could easily introduce patches and updates to mitigate the IoT malware propagation or attacks in a timely manner.
\item The future cyber security solution for IoT should take into the consideration that adversaries might leverage IoT devices with unpatched vulnerabilities to propagate malware via device-to-device links. In other words, the security mechanisms developed for IoT shall coexist with insecure, unpatched legacy IoT devices with uncontrolled device-to-device channels. A notification mechanism is suggested to help users identify the IoT devices at risk and further deny possible device-to-device connections. 
\end{itemize}

\section*{Acknowledgment}
This work was supported in part by Taiwan Information Security Center (TWISC), Academia Sinica, and Ministry of Science and Technology, Taiwan, under the grant MOST 104-2923-E-011-006-MY2, 105-2221-E-002-146-MY2, and 105-2218-E-001-001. 



\end{document}